\documentclass{ws-ijmpa}
\begin{document}
\markboth{Matt Visser and Nicolas Yunes}
{Power laws, scale invariance, and generalized Frobenius series...}
\catchline{}{}{}

\title{Power laws, scale invariance, and generalized Frobenius series:
  Applications to Newtonian and TOV stars near criticality}

\author{\footnotesize MATT VISSER}
\address{School of Mathematical and Computing Sciences, 
Victoria University of Wellington, \\ 
P.O. Box 600, Wellington,  New Zealand}

\author{NICOLAS YUNES}
\address{Washington University Gravity Group, Department of Physics, 
\\
Washington University in 
Saint Louis, Saint Louis, Missouri 63130, USA}

\maketitle
\pub{Received: 1 November 2002}{arXiv: gr-qc/0211001}
%
%
\def\d{{\mathrm{d}}}
\def\implies{\Rightarrow}
\def\be{\begin{equation}}
\def\bd{\begin{equation}}
\def\ba{\begin{eqnarray}}
\def\bea{\begin{eqnarray}}
\def\ee{\end{equation}}
\def\ed{\end{equation}}
\def\ea{\end{eqnarray}}
\def\eea{\end{eqnarray}}
\def\ie{{\emph{i.e.}}}
\def\eg{{\emph{e.g.}}}

\def\N{I\!\!N}
%
\begin{abstract}
%
  We present a self-contained formalism for calculating the background
  solution, the linearized solutions, and a class of generalized
  Frobenius solutions to a system of scale invariant differential
  equations.
  
  We first cast the scale invariant model into its equidimensional
  and autonomous forms, find its fixed points, and then obtain
  power-law background solutions.  After linearizing about these fixed
  points, we find a second linearized solution, which provides a
  \emph{distinct} collection of power laws characterizing the
  deviations from the fixed point. We prove that generically there
  will be a region surrounding the fixed point in which the complete
  general solution can be represented as a generalized Frobenius-like
  power series with exponents that are integer multiples of the
  exponents arising in the linearized problem. This Frobenius-like
  series can be viewed as a variant of Liapunov's expansion theorem.
  
  As specific examples we apply these ideas to Newtonian and
  relativistic isothermal stars and demonstrate (both numerically and
  analytically) that the solution exhibits oscillatory power-law
  behaviour as the star approaches the point of collapse.  These
  series solutions extend classical results; as exemplified for
  instance by the work of Lane, Emden, and Chandrasekhar in the
  Newtonian case, and that of Harrison, Thorne, Wakano, and Wheeler in
  the relativistic case.
  
  We also indicate how to extend these ideas to situations where fixed
  points may not exist --- either due to ``monotone'' flow or due to
  the presence of limit cycles. Monotone flow generically leads to
  logarithmic deviations from scaling, while limit cycles generally
  lead to discrete self-similar solutions.

\end{abstract}
%
\section{Introduction}
\label{C:Intro}
%
The presence of power-law behaviour in nature is such an extremely
common phenomenon that considerable lore has now grown up concerning
its genesis. One of the most common situations in which it occurs is
in the presence of scale-invariant systems. Such behaviour occurs, for
instance, in any sort of thermodynamic system undergoing a
second-order phase transition, and the behaviour of physical quantities
(such as susceptibilities) in terms of the distance from criticality
is typically given by a power-law:
\be \chi \propto
\left(T-T_\mathrm{critical}\over T_\mathrm{critical}\right)^\delta.
\ee 
Second-order phase transitions are extremely well-described by
statistical field theories.  The associated technical machinery (the
renormalization group), is now so well developed that it is sometimes
difficult to remember that second-order phase transitions are not the
\emph{only} route to power-law behaviour.  The onset of power-law
behaviour actually occurs at a much more primitive level, and can be
analyzed directly in terms of the underlying differential equations.

Scale invariance is commonly thought to be synonymous with power-law
behaviour. This is not quite correct: Scale invariance is not enough
to guarantee power-law behaviour, while power-law behaviour implies
scale invariance only in the trivial sense that it is always possible
to write down \emph{some} scale-invariant differential equation. (But
such a DE does not necessarily have to describe a physical theory.) In
the present paper, we examine scale invariant theories and find one
direct route to power-law behaviour.

This formalism consists of transforming a scale invariant theory into
its associated autonomous form\cite{Bender} and then determining the
location of its fixed points. These fixed points will lead inexorably
to a background solution that exhibits power-law behaviour.  By
expanding the system about its fixed point, a linearized autonomous
differential equation emerges, whose solution will exhibit
\emph{different} power-law behaviour.  This naturally leads to two
independent power laws: one for the background fixed point and a
second one for the linearized deviations from the fixed point.
Combining both results, our formalism indicates that the complete
general solution to the system will, over some limited region, have a
Frobenius-like power-series expansion. The usual Frobenius series is
one of the form
\be
y(x) = x^p \cdot \sum_{i=0}^\infty a_i \; x^i.
\ee
with $i$ a natural number, and is suitable for exploring linear ODEs
in the region of regular singular points. In the present context,
involving nonlinear DEs, the generalized Frobenius-like series we
encounter are generically of the form
\be
y(x) = x^p \cdot \sum_{i=0}^\infty  a_{i} \; x^{i \lambda}.
\ee
The exponent in the prefactor of this Frobenius-like series ($p$) is
governed by the exponent associated with the fixed-point power-law
background behaviour, while the power series is given in terms of
exponents ($i \lambda$) that are integer multiples of those arising in
the linearized problem ($\lambda$). This type of expansion is
essentially a modification of that given by Liapunov's expansion
theorem for autonomous DEs\cite{Lefschetz}, wherein we have now
back-tracked to give the expansion in terms of the original
scale-invariant system.

As specific examples we apply this formalism to both Newtonian and
relativistic isothermal stars, where the relevant differential
equations are equivalent to the exponential [isothermal] Lane--Emden
equation and its relativistic generalization. The resultant
Frobenius-like power series generalize the approximate solutions
found, for instance, by Chandrasekhar\cite{Chandrasekhar} and by
Harrison \emph{et.~al.}\cite{Harrison}, and provides significantly
greater analytic information concerning their structure.  This
formalism is also applied to a more general star, where the equation
of state need not be linear except in the core.  Both Newtonian and
relativistic stars are verified to undergo damped oscillations in the
total mass, pressure and radius as the central density goes to
infinity. These oscillations can accurately be described by a
Frobenius-like power series with complex exponents.  The relevant
exponents (though not the numerical values of the coefficients) drop
out naturally from the linearized problem.  This power law solution
matches the behaviour observed in numerical calculations to great
accuracy.
%
\section{Why power laws are ubiquitous}
\label{C:theory}
%
In this section we lay down the basic foundations of the formalism.
First, we define ``scale invariant'', ``equidimensional'', and
``autonomous'', and present the appropriate substitutions to go from
one form to another.  A single $n$th-order differential equation is
chosen as a model because of its simplicity, and then the formalism is
extended to a system of $I$ coupled $n$th-order differential
equations. After determining the existence of fixed points in the
equation, the background solution is obtained.  The differential
equation is then expanded about its fixed points to obtain a second
power-series solution near the fixed point (linearized solution).
Combining both results, a Frobenius-like series solution is obtained
in a neighborhood close enough to the fixed point.
%
\subsection{Fundamental definitions}
%
Let us define scale invariant differential equations in the usual
manner, as those which remain unchanged under the substitutions $x \to
a x$ and $y(x) \to a^p \; y(ax)$, where $a^p$ is a fixed but arbitrary
parameter.\cite{Bender}  Equations that remain unchanged under the
substitutions $x \to bx$ and $y(x) \to y(x)$, where $b$ is another
fixed but arbitrary parameter\cite{Bender}, will be defined as
equidimensional-in-$x$. Thus equidimensional equations can always be
viewed as scale invariant equations corresponding to the particular
value $p=0$ for the exponent.  Finally, autonomous equations will be
defined as those equations which remain unchanged under the
substitutions $x \to x + c$ and $y(x) \to y(x)$, where $c$ is a third
fixed but arbitrary parameter\cite{Bender}. This last condition is
that of translation invariance in the independent variable $x$.

It is a standard result that it is always possible to transform scale
invariant equations into equidimensional-in-$x$ equations through the
following substitution\cite{Bender}
\be
\label{Eq:1} 
y(x) = x^p \cdot w(x).
\ee
It is also a standard result that we can transform equidimensional-in-$x$
equations into autonomous equations by substituting\cite{Bender}
\be
\label{Eq:2}
x = e^t \qquad z(t) = w(x),
\ee
where $t$ is the new variable. 

%
\subsection{From scale invariance to autonomy}
%
Let us first consider a $n$th-order scale invariant differential
equation in a single dependent variable,
\be
\label{scale}
F \left(x, y(x), y'(x), y''(x),\ldots, {\d^n y(x)\over\d x^n} \right) = 0,
\ee
where the primes stand for differentiation with respect to $x$.  We
will write this as $F(x,y(x))=0$, but the presence of an appropriate
number of derivatives should be inferred. Now assuming this equation
to be scale invariant, let's apply equation (\ref{Eq:1}) to make
equation (\ref{scale}) equidimensional-in-$x$. The derivatives of
$y(x)$ become
\bd
{\d^n y(x)\over\d x^n} = \left({\d \over\d x}\right)^n \left[x^p \; w(x) \right] = 
\sum_{j=0}^n {{n! \; p!}\over {j! \left(n-j\right)! \left(p-j\right)!}} 
\; x^{p-j} \; w^{(n-j)}(x).
\ed
Note that the $j$th derivative of $y(x)$ involves only the $j$th and
lower-order derivatives of $w(x)$.  Rewriting $F$, we can define a new
function $\widetilde F$ as follows,
\begin{eqnarray}
&&
\widetilde {F} \left(x, w(x), w'(x),\ldots, {\d^n w(x)\over\d x^n} \right)
\equiv 
\\
&& 
\;
F \left(x, x^p w(x),p x^{p-1} w(x) +  x^p w'(x),\ldots, \sum_{j=0}^n 
{{n! p!}\over {j! \left(n-j\right)! \left(p-j\right)!}} x^{p-j} w^{(n-j)}(x) \right), 
\nonumber
\end{eqnarray}
since $F$ depends only on $x$, $w(x)$ and its derivatives.  The new
differential equation $\widetilde F(x,w(x))=0$ is now
equidimensional-in-$x$.

Let's now apply the substitution in equation (\ref{Eq:2}) to transform
it into an autonomous equation. The derivatives of $z$ are
\be
\label{Eq:3}
{\d^n \over\d x^n} = \left(e^{-t} {\d \over\d t}\right)^n = 
e^{-n t} \prod_{j=0}^{n-1} \left({\d \over\d t} - j \right).
\ee
As before, note here that the product in equation (\ref{Eq:3})
involves an $n$th order term plus lower-order derivatives (and no
higher-order derivatives). Rewriting $\widetilde F$ we can define a
new function $\bar F$ via
\bea
&&\bar F \left(z(t), \dot{z}(t),\ldots, {\d^n \over\d t^n} z(t) \right) \; f(t) 
= 
\nonumber\\
&&\qquad 
\widetilde {F} \left(e^t, z(t),  e^{-t}\dot{z}(t),\ldots, e^{-n t} \prod_{j=0}^{n-1} 
\left({\d \over\d t} - j \right) z(t) \right),
\eea
where the dots denote differentiation with respect to $t$ and $f(t)$
is some arbitrary function of $t$.  The function $\bar F(z(t))$ is now
translation invariant, since the substitution $t \to t + b$ as $z(t)
\to z(t + b)$ leaves the equation $\widetilde F(x,w(x))=0$ (based on
setting the RHS $=$ zero) unchanged. The only way this can be true is
if we can factorize $\widetilde F(x,w(x))$ into an autonomous $\bar F$
times some $f(t)$, yielding the LHS.

Our original scale invariant equation $F(x,y(x))=0$, that had become
equivalent to the equidimensional-in-$x$ equation $\widetilde
F(x,w(x))=0$, has now become equivalent to the autonomous equation
$\bar F(z(t))=0$.  We can of course also work backward and prove that
for any autonomous-in-$t$ equation for the dependent variable $z(t)$
the substitution $x=e^t$ (that is, $\d/\d t = x \,\d/\d x$) produces an
equation that is equidimensional-in-$x$ with $z(t) = z(\ln x) = w(x)$.
Furthermore for any value of $p$ the substitution $w(x) = x^{-p}\, y(p)$
results in a differential equation that is scale invariant. However,
this scale invariant equation need not be equal to the physical
differential equation that describes the system we are concerned with.
Without some additional information beyond the mere existence of a
power law, there is simply not much one can say about the physics.
%
\subsection{Scale-invariant ordinary differential equations of $n$th-order}
\subsubsection{Background solution}
%
In order to obtain the background solution of the ODE, we must look at
the fixed points. Let's assume that $\bar F$ has a fixed point $z_*$
such that all derivatives of $z(t)$ vanish at that fixed point, \ie,
\be 
\bar {F}\left( z(t) = z_*, {\d^i z(t)\over\d t^i} = 0 \right) = 0, 
\ee
where $i\in[1,n]$ denotes any positive integer in that range.  Hence,
a solution to the differential equation $\bar {F}(z(t)) = 0$ is $z(t)
= z_*$. Tracing back all the substitutions we made, we obtain
\be
\label{Eq:4}
y(x) = x^p \; z_* \qquad\mbox{which solves}\qquad {F}(x,y(x)) = 0.
\ee
From this equation, the power law behaviour is obvious. However, not
all scale invariant equations possess fixed points. A scale invariant
differential equation, whose associated autonomous equation \emph{does
  not} have any fixed points, \emph{will not} possess power-law
solutions. Refer to \ref{appe:1} for an explicit counter-example.  In
light of this, we can formulate the following theorem:
\begin{theorem} 
  Any first-order differential equation that is scale invariant
  \emph{and} whose associated autonomous equation possesses a fixed
  point will have a \emph{background} solution in the form of a power
  law [equation (\ref{Eq:4})] with exponent given by the scale
  invariant condition.
\end{theorem}
%
\subsubsection {Linearized solution}
%
Let us linearize the autonomous equation by setting $z(t) = z_* +
z_1(t) + O[z_1(t)^2]$, where $z_*$ is the fixed point and $z_1(t)$ is
a small perturbation about $z_*$.  $\bar F(z(t))=0$ reduces to
\be
\bar F(z(t)) = \bar F(z_*) + \bar F_1(z_1(t)) +  O[z_1(t)^2] =  
\bar F_1(z_1(t)) +  O[z_1(t)^2] = 0.
\ee
So in the linear approximation, we must have
\be
\bar F_1(z_1(t)) = 0.
\ee
But $\bar F_1(z_1(t))=0$ is an ODE which is both autonomous and
linear, and still retains its character as a $n$th-order differential
equation. Autonomous plus linear is a very powerful constraint, and in
fact implies a differential equation with constant coefficients. That
is, there is some polynomial $P(\cdot)$ such that the function $\bar
F_1(z_1(t))$ satisfies
\be
\bar F_1(z_1(t)) \equiv P\left({\d\over\d t}\right) z_1(t).
\ee
The standard technique\cite{Bender} for solving differential equations
of this type is to insert a trial solution of the form $z_1(t) = A
\exp(\lambda t)$. Then the differential equation reduces to the simple
$n$th-order polynomial
\be P(\lambda) = 0, \ee 
which has at most $n$ distinct solutions.  Assuming for the time being
that all roots are distinct, the general solution to the linearized
autonomous equation is
\be
z_1(t) = \sum_{i=1}^n A_i \; \exp(\lambda_i t).
\ee
That is
\be
z(t) = z_* + \sum_{i=1}^n A_i \; \exp(\lambda_i t) + O(A^2).
\ee
In terms of the original variable $y(x)$ 
\be
\label{theo:1}
y(x) = x^p \left[ z_* + \sum_{i=1}^n A_i \; x^{\lambda_i} + O(A^2) \right].
\ee
Near the fixed point, we now see the presence of multiple power laws.
The prefactor $x^p$ is governed directly by the scale invariant
condition, while the subsidiary exponents arise from solving a
polynomial equation based on the linearized approximation.

We are now in a position to formulate the following theorem:
\begin{theorem} 
\label{theo:5}
Any $n$th-order ordinary differential equation that is scale invariant
\emph{and} whose associated autonomous equation possesses a fixed
point will have a linearized solution which is generically expressible
in the form of a power law (equation \ref{theo:1}) in a neighborhood
surrounding the fixed point.
\end{theorem}

We can also enunciate a general rule of thumb: Because the exponent
$p$ ultimately arises from algebraic manipulations involving the
physical dimensionalities of the various quantities present in the ODE
it is likely to be a rational number. In contrast the exponents
$\lambda_i$, being solutions of a polynomial equation, will
generically be irrational.

In the exceptional case where the roots of the polynomial $P(\lambda)$
are not distinct, then there are additional technical complications.
Let $m$ of the roots be distinct, and let $g_i$ with $i\in[1,m]$ be the
multiplicity of the $i$th distinct root. Then we have $\sum_i^m g_i =
n$ and
\be
z_1(t) = \sum_{i=1}^m   \sum_{j=1}^{g_i-1} A_{ij} \; t^{i-1} \; \exp(\lambda_i t).
\ee
That is
\be
z(t) = z_* + \sum_{i=1}^m   \sum_{j=1}^{g_i-1} A_{ij} \; t^{i-1} \; \exp(\lambda_i t) + O(A^2).
\ee
In terms of the original variable $y(x)$ 
\be
\label{theo:6}
\label{E:logs}
y(x) = x^p \left[ z_* +   \sum_{i=1}^m \sum_{j=1}^{g_i-1} A_{ij} \; (\ln x)^{i-1} \; x^\lambda_i + O(A^2) \right].
\ee
Thus logarithmic deviations from power law behaviour, while not
generic, are certainly relatively easy to get --- despite the lore
based in statistical field theory and second-order phase transitions a
logarithm is not necessarily the result of one-loop physics. Logarithms
can arise from causes as mundane as a repeated root in a linearized
autonomous ODE.

%
\subsubsection{Frobenius-like solution}
%
Going beyond the linearized approximation,  in the vicinity of
the fixed point, one can write the expansion more systematically as
$z(t) = z_* + \epsilon z_1(t) + \epsilon^2 z_2(t) + O[\epsilon^3]$, 
where $\epsilon \ll 1$, in which case $\bar F(z(t))=0$ reduces to
\be
\bar F(z(t)) = \bar F(z_*) + \epsilon \bar F_1((z_1(t)) + 
\epsilon^2 \left[ \bar F_1(z_2(t)) +  \bar F_2(z_1(t),z_1(t)) \right] + O(\epsilon^3).
\ee
Here $\bar F_2(\cdot,\cdot)$ represents a second-order functional
derivative of $\bar F(\cdot)$ around the fixed point, which will be
some messy quadratic differential operator with constant coefficients.
This implies both
\be
\bar F_1(z_1(t)) = 0,
\ee
and 
\be
\bar F_1(z_2(t)) = -  \bar F_2(z_1(t),z_1(t)).
\ee
That is
\be
z_2(t) = -  \bar F_1{}^{-1}\left[\bar F_2(z_1(t),z_1(t))\right].
\ee
But we already know that $z_1 = \sum_{i=1}^n A_i \;
\exp(\lambda_i t)$, so $z_2$ consists of pieces of the form
$\exp([\lambda_i+\lambda_j] t)$. Indeed
\be
z_2(t) = \sum_{i=1}^n \sum_{j=1}^n B(A_i,A_j) \; \exp([\lambda_i+\lambda_j] t).
\ee
with $B(A_i,A_j)$ being some complicated function of the $A_i$ that is
deducible from the precise form of $\bar F_1$ and $\bar F_2$. 

Collecting terms
\be
z(t) = z_* + \sum_{i=1}^n A_i \; \exp(\lambda_i t) +  
\sum_{i=1}^n \sum_{j=1}^n B(A_i,A_j) \; \exp([\lambda_i+\lambda_j] t) +   O(\epsilon^3).
\ee
In terms of the original variable $y(x)$ 
\be
y(x) = x^p \cdot \left[ z_* + \sum_{i=1}^n A_i \; x^{\lambda_i} + 
\sum_{i=1}^n \sum_{j=1}^n B(A_i,A_j) \; x^{\lambda_i+\lambda_j} +
O(\epsilon^3) \right].
\ee
The general pattern is now clear. Let us take $\vec \lambda =
(\lambda_1,\lambda_2, \dots \lambda_n)$ and let $\vec m =
(m_1,m_2,\dots m_n) \in \N^n$. Then formally
\be
z(t) = \sum_{\vec m \in \N^n} A(\vec m) \; 
\exp\left(\vec \lambda \cdot \vec m \; t\right).
\ee
This series may be asymptotic, in general we make no claims about
convergence, though in specific cases it often does converge in an
open region surrounding the fixed point. The coefficients $A(\vec m)$
are certainly not independent but are inter-related via the underlying
differential equation. This is basically Liapunov's expansion theorem,
as discussed for instance by Lefschetz\cite{Lefschetz}. Readers
interested in additional mathematical rigour are directed to that
reference.

Near the fixed point our equation becomes, in terms of the original
variables $y(x)$,
\be
\label{Theo:2}
y(x) = x^p \cdot \sum_{\vec m \in \N^n} A(\vec m) \; x^{\vec \lambda \cdot \vec m}.
\ee
This type of series is Frobenius-like in the sense that the prefactor
$x^p$ will be governed by the background solution, while the exponents
of the power series will be given by the linearized problem.  Hence,
we can formulate the following theorem:
\begin{theorem} 
  Any $n$th-order ordinary differential equation that is scale
  invariant \emph{and} whose associated autonomous equation possesses
  a fixed point will generically have a formal solution in the form of
  a Frobenius-like power series (equation (\ref{Theo:2})) in a
  neighborhood close to the fixed point.
\end{theorem}

As per the previous discussion, this form holds only if the roots of
the linearized problem are distinct. For coincident roots one should
expect the Frobenius-like series above to be modified by various
logarithmic terms. Indeed based on Liapunov's expansion
theorem\cite{Lefschetz} the coefficients $A(\vec m)$ become polynomial
in $t$, and so polynomial in $\ln z$ [\emph{cf.} equation
(\ref{E:logs}) above].  The resulting series is too clumsy to be of
immediate use and fortunately none of the specific examples we
encounter exhibit this particular behaviour.

%
\subsection{Systems of $n$th-order scale-invariant differential equations}
\subsubsection{Background solution}
%
All of the previous analysis can be generalized to systems of $I$
coupled $n$th-order ordinary differential equations.  There are three possible routes:
\begin{enumerate}
\item Develop a suitable ``vector'' and ``matrix'' notation.
\item Use the fact that a system of $I$ coupled $n$th-order
  differential equations can always be reduced to a system of $N =
  I\times n$ coupled first-order differential equations\cite{Bender},
  but you would still need to develop a suitable ``vector'' and
  ``matrix'' notation.
\item Use the fact that a system of $I$ coupled $n$th-order
  differential equations can generically be ``decoupled'' by being
  reduced to a single DE of order $N = I\times n$ in one of the
  dependent variables, though there seems to be no constructive
  algorithm for doing so.\cite{Bender}
\end{enumerate}
On balance, we have found it most useful to directly develop a
suitable ``vector'' and ``matrix'' notation.  Let us begin by
considering the equation 
\be \left[F\right] = 0 \ee 
where $\left[F\right]$ is the column vector of ODEs
\bd 
\left[F\right] = 
  \left(
  \begin{array}{c}
    F_1\left(x, y_1(x),y_1'(x),\ldots, y_2(x),y_2'(x),\ldots, y_I(x),
             \ldots,{\d^n {y_I}(x)\over\d x^n} \right) \\
    F_2\left(x, y_1(x),y_1'(x),\ldots, y_2(x),y_2'(x),\ldots, y_I(x),
             \ldots,{\d^n {y_I}(x)\over\d x^n} \right) \\
    \vdots \\
    F_I\left(x, y_1(x),y_1'(x),\ldots, y_2(x),y_2'(x),\ldots, y_I(x),
             \ldots,{\d^n {y_I}(x)\over\d x^n} \right) \\
                 \end{array} \right).
\ed
When we say that this system is scale invariant, we mean that as $x
\to ax$ and
\be
\left[Y(x)\right] \to \left[A\right] \left[Y(x)\right],
\ee
where
\bd 
\left[A\right] = \left[ \begin{array}{ccccc} 
                 a^{p_1} & 0 & 0 & \ldots & 0 \\
                 0 & a^{p_2} & 0 & \ldots & 0 \\
                 \vdots & \vdots & \vdots & \ddots & \vdots \\
                 0 & 0 & 0 & \ldots & a^{p_I} \\
                 \end{array} \right]
\ed
and $\left[Y(x)\right]$ is a column vector of dependent variables, the
equation $\left[F\right] = 0$ remains invariant. Note that this means
that we have chosen our dependent variables $y_i(x)\dots y_I(x)$ in
some convenient manner to make their scale transformation properties
both simple and independent of each other --- this will almost always
occur automatically in systems of ODEs derived from an underlying
physical problem.

Previously, we also showed that any $n$th-order differential equation
in a single dependent variable that is scale invariant can be put into
its associated autonomous form by applying the substitutions described
in equations (\ref{Eq:1}) and (\ref{Eq:2}). Therefore, it carries over
that we can make the same kind of substitution in each equation of the
system. After doing so, we can define a new column vector $\left[\bar
  F\right]$ in the following manner:
\bd
\left[\bar {F}\right] = \left( \begin{array}{c}
                    \bar {F_1}\left(z_1(t),\dot{z_1}(t),\ldots, z_2(t),\dot{z_2}(t),
                              \ldots, z_I(t),\ldots,{\d^n {z_I}(t)\over\d t^n} \right) \\
                    \bar {F_2}\left(z_1(t),\dot{z_1}(t),\ldots, z_2(t),\dot{z_2}(t),
                              \ldots, z_I(t),\ldots,{\d^n {z_I}(t)\over\d t^n} \right) \\
                    \vdots \\
                    \bar {F_I}\left(z_1(t),\dot{z_1}(t),\ldots, z_2(t),\dot{z_2}(t),
                              \ldots, z_I(t),\ldots,{\d^n {z_I}(t)\over\d t^n} \right) \\
                   \end{array} \right),
\ed
where for each $i\in[1,I]$ we have $y_i(x) = x^{p_i} \; w_i(x) =
e^{tp_i} \; z_i(t)$ and furthermore $x = e^t$. From the previous
analysis we know that each equation in this new vector equation
$\left[ \bar {F} \right]=0$ is now autonomous, hence, the whole system
is autonomous. Let's now assume that this autonomous system has a
fixed point $\left[Z^*\right]$ such that all derivatives of $z(t)$
with respect to $t$ in $\left[ \bar {F} \right]$ vanish at the fixed
point, where we mean
\bd
\left[Z^*\right] = \left( \begin{array}{c} z_1^* \\ z_2^* \\  \vdots \\ z_I^* \end{array} \right).
\ed

That is 
\be
\left[ Z(t) \right] 
=
\left(\begin{array}{c} z_1(t) \\ z_2(t) \\ \vdots \\ z_I(t) \end{array}\right)
=
\left[Z^*\right]
=
\left(\begin{array}{c} z_1^* \\ z_2^* \\ \vdots \\ z_I^* \end{array}\right).
\ee
is one specific solution to the system of ODEs.
 
Substituting back we obtain the following equations for $\left[Y(x)\right]$:
\be
\label{Theo:3}
\left(\begin{array}{c} y_1(x) \\ y_2(x) \\ \vdots \\ y_I(x) \end{array} \right) = 
\left[\begin{array}{ccccc} x^{p_1} & 0 & 0 & \ldots & 0 \\
                     0 & x^{p_2} & 0 & \ldots & 0 \\
                     \vdots & \vdots & \vdots & \ddots & \vdots \\
                     0 & 0 & 0 & \ldots & x^{p_I} \\
\end{array} \right] 
\left(\begin{array}{c} z_1^* \\ z_2^*  \\ \vdots \\ z_I^* \end{array} \right) 
=
\left(\begin{array}{c} z_1^* \; x^{p_1} \\ z_2^*  \; x^{p_2}\\  
      \vdots \\ z_I^*  \; x^{p_I}\end{array} \right).
\ee

Therefore, we can formulate the following theorem:
\begin{theorem}
\label{Thm:1}
Any system of $I$ $n$th-order differential equations that is scale
invariant, \emph{and} whose associated system of autonomous equations
possesses a fixed point, will have solutions in the form of a
collection of power laws (equation \ref{Theo:3}), with exponents given
by the scale invariant condition.
\end{theorem}
%
\subsubsection{Linearized solution}
%
If we linearize the $n$th order system, then for each $\bar F_i$ we
will obtain a linear differential equation that has the capability of
mixing the various $z_i$. Specifically let
\be
z_i(t) = z^*_i + \epsilon z_{1,i}(t) + O(\epsilon^2),
\ee
then
\be
\bar F_i(z_j(t)) = \bar F_i(z^*_j) + \sum_{j=1}^n \{\bar F_1\}_{ij}(z_{1,j}) 
+  O(\epsilon^2).
\ee
Here the $\{\bar F_1\}_{ij}$ are linear differential operators that
lead to a system of linear and autonomous differential equations
\be
\sum_{j=1}^n \{\bar F_1\}_{ij}(z_{1,j}(t)) = 0.
\ee
This implies that each of the $\{\bar F_1\}_{ij}$ is a constant
coefficient polynomial in $\d/\d t$
\be
\{\bar F_1\}_{ij} \equiv P_{ij}\left({\d\over\d t}\right).
\ee
Inserting a trial solution of the form $z_{1,j}(t) = A_j \exp(\lambda
t)$ now yields the vector equation
\be
\sum_{j=1}^n  P_{ij}(\lambda) \; A_j = 0.
\ee
There is a non-trivial solution if and only if the $I\times I$ matrix
$P_{ij}(\lambda)$ is singular ($\det\{P_{ij}(\lambda)\} =0$), in which
case the $A_i$ correspond to the singular eigenvector. Since each
element of the matrix $P_{ij}(\lambda)$ is itself a polynomial of
order $n$, (where $n$ is the order of the individual differential
equations in the system), the determinant is a polynomial of order $N
= I \times n$ having up to $N$ separate roots. Let us assume these
roots are distinct and call them $\lambda_\alpha$ with
$\alpha\in(1,N)$ and denote the corresponding eigenvector by
$(A_\alpha)_i$. As long as the roots are distinct these eigenvectors,
multiplied by the associated exponentials, will span the solution
space and we can write the general solution to our autonomous system
as
\be
z_{1,i}(t) = \sum_{\alpha=1}^N (A_{\alpha})_i \; \exp(\lambda_\alpha t).
\ee
In terms of our original equation,
\be
\label{Theo:4}
y_i(x) = x^{p_i} \left\{ z^*_i + \sum_{\alpha=1}^N (A_\alpha)_i \; 
x^{\lambda_\alpha} +O(\epsilon^2) \right\}.
\ee
This now allows us to formulate the following theorem:
\begin{theorem} 
  Any system of $I$ $n$th-order differential equations that is scale
  invariant \emph{and} whose associated autonomous equation possesses
  a fixed point will generically have solutions in the form of a
  collection of power laws (equation \ref{Theo:4}) in a neighborhood
  close to the fixed point.  The prefactors $[x^{\mathcal P}]$ will be
  governed by the background solution, while the $\lambda_\alpha$ are
  given by the linearized problem.
\end{theorem}

If the roots are not distinct then a much messier equation involving
logarithms in $x$ can be constructed, but is beyond the scope of the
applications we have in mind.

%
\subsubsection{Frobenius-like solution}
\label{Frob}
We can again go beyond the linearized approximation in the vicinity of
the fixed point and write $z_i(t) = z_i^* + \epsilon z_{1,i}(t) +
\epsilon^2 z_{2,i}(t) + O[\epsilon^3]$. Then,
\bea
&&\bar F_i(z_j(t)) = \bar F_i(z_j^*) 
+ \epsilon \sum_{j=1}^n \{\bar F_1\}_{ij}(z_{1,j}) 
\nonumber
\\
&&
\qquad
+ \epsilon^2 \left[ \sum_{j=1}^N \{\bar F_1\}_{ij}(z_{2,j}) +  
\sum_{j=1}^N\sum_{k=1}^N\{\bar F_2\}_{ijk}(z_{1,j}(t),z_{1,k}(t)) \right] 
+ O(\epsilon^3),
\eea
where, similarly to the preceding discussion for a single dependent
variable, the quantity $\{\bar F_2\}_{ijk}(\cdot,\cdot)$ represents
the second-order functional derivative of $\{\bar F\}_{i}(\cdot)$
around the fixed point $[Z^*]$.  Using the same reasoning as before,
we will obtain a solution of the form
\be
z_{2,i}(t) = \sum_{\alpha=1}^N \sum_{\beta=1}^N 
\{B(A_\alpha,A_\beta)\}_i \; \exp([\lambda_\alpha+\lambda_\beta] t).
\ee
Collecting terms and substituting back to our original solutions of
$y_i(x)$ we obtain
\be
y_i(x) = x^{p_i} \left[ z_i^* + \sum_{\alpha=1}^N \{A_\alpha\}_i 
      \; x^{\lambda_\alpha} + 
\sum_{\alpha=1}^N \sum_{\beta=1}^N 
\{B(A_\alpha,A_\beta)\}_i \; x^{\lambda_\alpha + \lambda_\beta} +
O(\epsilon^3) \right],
\ee
The general pattern is again clear
\be
\label{Theo:6}
y_i(x) = x^{p_i} \cdot \sum_{\vec m \in \N^N} \{A(\vec m)\}_i \;\;
 x^{\vec \lambda \cdot \vec m},
\ee
where as before, $\vec \lambda = (\lambda_1,\lambda_2, \dots \lambda_N)$ and 
$\vec m = (m_1,m_2,\dots m_N) \in \N^{N}$.
We can now formulate the following theorem:
\begin{theorem} 
  Any system of $I$ $n$th-order differential equations that is scale
  invariant \emph{and} whose associated autonomous equation possesses
  fixed points $z_{i*}$ will generically have a solution in the form
  of a Frobenius-like power series (equation \ref{Theo:6}) in a
  neighborhood close to the fixed point.  The prefactors $x^{p_i}$
  will be governed by the background solution, while the exponents of
  the power series will be given by the linearized problem.
\end{theorem}
As previously, there is an explicit requirement that the exponents
arising from the linearized problem be distinct, otherwise (as per
Liapunov's expansion theorem) messy expressions polynomial in $\log x$
will be encountered. (See Lefschetz\cite{Lefschetz} for technical
details.)
%
%

\subsection{Discussion}
\label{SS:generic-discussion}

Before we turn to several specific astrophysical examples to show how
this formalism works in detail, let us summarize key points:
\begin{itemize}
\item Scale invariance does not automatically imply power-law
  behaviour; the existence of one or more fixed points is an
  additional requirement. An explicit example of a scale-invariant ODE
  without any power-law solution is presented in \ref{appe:1}.
\item If we perturb slightly away from the fixed points, there will no
  longer be an exact power-law behaviour, but we have something
  similar: Generically there will be a Frobenius-like expansion of the
  form
\be
y(x) = x^p \cdot \sum_{i=0}^\infty  a_{i} \; x^{i \lambda},
\ee
where the exponent $p$ depends on the fixed point and the exponent
$\lambda$ depends on the linearized ODE around the fixed point.  The
coefficients $a_i$ are of course constrained by the original exact
ODE, and since it is $n$th-order we expect $n$ of these coefficients
to be ``free'' (that is, to depend on initial conditions) with the
other coefficients in principle being determinable from the first $n$
and the exact ODE.
\item We feel that this formalism, attributing as it does power-law
  behaviour to rather general features of differential equations, goes
  a long way towards explaining the apparent ubiquitous occurrence of
  power-laws in nature.
\end{itemize}

\section{Application: Isothermal Newtonian stars}
\label{C:newton}
In the following section, we apply the general formalism developed
above to a static isothermal Newtonian star. The relevant system of
ODEs is scale invariant, and contains two fixed points, one
corresponding to the center of the star, and the other corresponding
to the point at infinity (where the star develops an infinitely thin
halo).  We formally solve the differential system associated with it
and obtain power-series solutions for the total pressure, radius,
compactness and mass of the star in terms of a Frobenius-like series
about either fixed point.  For each of these physical quantities, the
radius of convergence of these series (one for each fixed point)
overlap, so they constitute a complete power-series solution.  These
results subsume and extend previous results found by
Chandrasekhar\cite{Chandrasekhar}, and the classic results of Lane and
Emden.  In particular, these series are then used to discuss the
behaviour of such stars near the onset of collapse.
\subsection{Background solution}
%
The basic equations of (static) non-relativistic stellar
structure\cite{Chandrasekhar} are ($G\equiv1$)
\bd
{\d p\over\d r} = - \rho(r) \; {m(r)\over r^2};
\ed
\bd
m(r) = \int_0^{r} 4\pi\;{r'}^2\;\rho(r')\; \d{r'}.
\ed
Let us now adopt the following particularly simple equation of state, which is
appropriate, for instance, for an isothermal star:
\be
p(r) = c_s^2 \; \rho.
\ee
Then
\be
{\d p\over\d r} = - {p(r)\over c_s^2} \; {m(r)\over r^2}, 
\ee
\be
\label{mass}
{\d m(r)\over\d r} = {4\pi p(r) r^2\over c_s^2}, 
\ee 
where $c_s$ stands for the speed of sound.  The relevant boundary
conditions are $m(0)=0$ and $p(0)=p_0$.  This system of differential
equations is scale invariant under $r \to a r$, $p \to a^{-2} p$ and
$m \to a^{+1} m$. This means that from our previous analysis we have
$(y_1,y_2)=(p,m)$ and $(p_1,p_2)=(-2,1)$.  As stated previously, we
can make suitable transformations to convert this scale invariant
system into an autonomous system.  For this purpose, let's substitute
\be
\label{zeta:3}
\zeta(r)=2\pi p(r) r^2/c_s^4,
\ee
and
\be
\label{chi:3}
\chi(r)=m(r)/\left(2 c_s^2 r \right).
\ee
Then, the system becomes
\be
\label{zeta:1}
r\; {\d \zeta(r)\over\d r} =  2 \zeta(r) \left[1-\chi(r)\right];
\ee
\be
\label{chi:1}
r\; {\d \chi(r)\over\d r} = \zeta(r) - \chi(r).
\ee
As expected this system is now equidimensional-in-$r$.  Define a new
variable $t=\ln(r)$ so $\d/\d t = r \d/\d r$, then
\be
{\d \zeta(t)\over\d t} =  2 \zeta(t)\; \left[1-\chi(t)\right];
\ee
\be
{\d \chi(t)\over\d t} = \zeta(t) - \chi(t).
\ee
As we can clearly see this system is now autonomous.  There is a fixed
point at $\chi=1$ and $\zeta=1$. For any initial condition this point
is an attractor as $t\to\infty$. The center of the star is at $r=0$
($t=-\infty$). For a regular star with finite $p_0$ we have $\zeta=0$
and $\chi=0$ as $t\to-\infty$; which is the second fixed point. So
solving our special dimensionless differential equation is equivalent
to looking for the unique integral curve that emerges from the fixed
point at $(0,0)$ and terminates on the fixed point at $(1,1)$.
Applying Theorem \ref{Thm:1} we can easily check that a singular
solution to the system is of the form
\be
p(r) = C_1 \; r^{-2},
\ee
and
\be
m(r) = C_2 \; r^1,
\ee
where $C_1$ and $C_2$ are constants. Solving for the constants we see
that the $(1,1)$ fixed point corresponds to the power-law solution
\be
\label{Eq:5}
m(r) = 2 \;c_s^2\; r,
\ee
and 
\be
\label{Eq:6}
p(r) = {c_s^4\over2\pi r^2},
\ee
while the $(0,0)$ fixed point corresponds to the more prosaic
$p(r)=0=m(r)$.  
%
\subsection{Power series in terms of the radius}
\subsubsection{Linearization about the fixed points}
%
In order to obtain the linearized solution it is necessary to
linearize about its critical points.  Let us first make the
substitution $\chi = \chi_1 + 1$ and $\zeta = \zeta_1 + 1$. Then,
\bd
{\d \zeta_1(t)\over\d t} =  - 2 \chi_1(t);
\ed
\bd
{\d \chi_1(t)\over\d t} = \zeta_1(t) - \chi_1(t).
\ed
Close enough to the $(1,1)$ stagnation point, solutions are of the
form
\be
{\zeta(t)\choose\chi(t)} = 
\exp\left\{ \left[ \begin{array}{cc}0\;&-2\\1\;&-1\end{array} \right] (t-t_i) \right\}
\; {\zeta(t_i)\choose\chi(t_i)}. 
\ee
The eigenvalues of this matrix are $\delta = (-1\pm i \sqrt{7})/2$,
and the approximate linearized solutions approach the critical point
exponentially in $t$ --- as $\exp(\delta \; t) = r^\delta$.

There is a second (unstable) stagnation point at $(0,0)$ --- a
repeller. If we linearize around $(0,0)$ we get
\be
{\d \zeta_1(t)\over\d t} =  +2 \zeta_1(t);
\ee
\be
{\d \chi_1(t)\over\d t} = \zeta_1(t) - \chi_1(t).
\ee
The corresponding eigenvalues are $(\gamma_1, \gamma_2) = (2,-1)$,
corresponding to regular $r^2$ behaviour for $\zeta$ and $\chi$ at the
center of the star, plus an otherwise possible but irregular and
unphysical behaviour for $\chi$ that goes as $1/r$. This $1/r$
behaviour would correspond to a finite $m(0)$ if such were possible.

\subsubsection{Power series about $(0,0)$ in terms of the radius}
Based on Theorem \ref{theo:5} and our general comments regarding
Frobenius-like series it is reasonable to assume the following
expansion to solve the system
\be
\label{Eq:8}
\chi(z) = \sum_{m=0}^\infty \sum_{n=0}^\infty b_{mn} \; z^{m\gamma_1+n\gamma_2}.
\ee
Here we have introduced the dimensionless variable
\be
z=\sqrt{4\pi\rho_0/c_s^2} \; r = (\sqrt{4\pi p_0}/c_s^2) \; r, 
\ee
with the particular numerical coefficients being chosen for future
convenience.  In this case, since the $\gamma_2 = -1$ index is
unphysical, it will be excluded from the analysis. Hence, we restrict
attention to an expansion of the following form,
\bd
\chi(z) = \sum_{n=0}^\infty b_{n} \; z^{n\gamma_1} \to \sum_{n=0}^\infty b_{n} \; z^{2n}.
\ed

In order to solve for the correct coefficients $b_{n}$ we need to plug
the previous equation into our differential equations. First, by
uncoupling the system (eliminating $\zeta$ to obtain a single
second-order differential equation for $\chi$) we obtain
\be
{\d^2 \chi(z)\over\d z^2} - {2\chi(z)\over  z^2} + {2\chi(z)^2\over z^2} +
{2\chi(z)\over z} \;{\d \chi(z)\over\d z} =0.
\ee
The relevant boundary conditions may be inferred from the fact that
the pressure (and so also the density) is finite at the origin:
\be
\chi(0)=\chi'(0)=0; \qquad \chi''(0) =  {1\over3}; \qquad \chi' \equiv \d\chi/\d z.
\ee
The fact that only integer exponents arise in our series solution for
$\chi(z)$ is ultimately due to the fact that the second-order ODE
above, though nonlinear, has what amounts to a regular fixed point at
$z=0$.  So standard results in terms of ordinary Frobenius series
suffice to develop an appropriate ansatz.

Now multiplying by $z^2$, we can write 
\be
\label{Eq:7}
\left(z{\d\over\d z}\right)^2 \chi -  z{d\over\d z}  \chi - 2 \chi 
= 
- 2 \chi^2 - 2\chi\;\; z{\d\over\d z} \chi,
\ee
or in operator form,
\be
\label{operator}
\left[ \left(z{\d\over\d z}\right)^2  -  z{d\over\d z}  - 2 \right] \chi 
= 
- \left[ 2 + z{\d\over\d z}\right] \chi^2.
\ee

Now we plug in our assumption of the expansion in equation (\ref{Eq:8}) 
and we get a recursion relation of the form 
\be
\left[(2n)^2 - (2n) -2\right] b_n = 
-\left[2 + (2n) \right]  \sum_{i=0}^n b_i \; b_{n-i}.
\ee
Because $b_0=0$ we can rewrite this as
\be
\label{E:recur1}
b_n = -{(n+1)\over(n-1)(2n+1)}  \sum_{i=1}^{n-1} b_i \; b_{n-i}.
\ee
This recursion relation, together with the initial condition $b_1=1/6$,
derived from the fact that $\chi''(0)=1/3$, completely specifies the
power series about the $(0,0)$ critical point. It is straightforward
to compute these coefficients numerically, though there does not seem
to be any simple explicit closed form for the general term.

For $\zeta$ we use the relation 
\be 
z {\d\over\d z} \chi = \zeta - \chi.
\ee 
So if 
\be 
\zeta(z) = \sum_{n=0}^\infty a_n z^{2n},
\ee 
then
\be 
a_n = (2n+1) \; b_n.  
\ee 
Keeping the first 250 terms in this series, and comparing with an
explicit $4$th order Runge--Kutta integration, gives very high
accuracy [$O(10^{-6})$] out to $z\approx3$.

In terms of the physical variables $m(r)$ and $p(r)$ we have
\be
m(r) = 2 c_s^2 r \chi(r) 
= 2 c_s^2 r \sum_{n=1}^\infty b_n \left({4\pi\rho_0 r^2\over c_s^2}\right)^n
= 2 c_s^2 r \sum_{n=1}^\infty b_n \left({4\pi p_0 r^2\over c_s^4}\right)^n,
\ee
and
\be
p(r) = {\zeta(z) c_s^4\over 2 \pi r^2} 
=   
{c_s^4\over 2 \pi r^2} \sum_{n=1}^\infty a_n \left({4\pi\rho_0 r^2\over c_s^2}\right)^n
=   
{c_s^4\over 2 \pi r^2} \sum_{n=1}^\infty a_n \left({4\pi p_0 r^2\over c_s^4}\right)^n.
\ee
These are Taylor series for the solution to the differential equations
that describe an isothermal star, with the coefficients being
specified by the recursion relations determined above. By bounding the
coefficients using the recursion relation, one can show that the
series definitely converges for $|z| \leq 3$ and that it definitely
diverges for $|z| \geq \sqrt{12}$. Numerical methods seem to indicate
a radius of convergence of $R \approx 3.273687274$.  The $a_n$ and
$b_n$ coefficients do not have a simple explicit analytic form, but
computing them numerically from the recursion relation is trivial.

Note that in more classical language this analysis is equivalent to
finding a power series solution of the isothermal Lane--Emden
equation. See \ref{a:lane-emden} for details.

%
\subsubsection{Power series about $(1,1)$ in terms of the radius}
%
Turning to the other critical point, from equation (\ref{Eq:7}) we see
that $\chi(\infty) = 1$.  Now making a shift of variables appropriate
to the $(1,1)$ fixed point, $\chi = 1+\chi_1$, where $\chi_1$ is
small, we obtain
\be
\label{Eq:9}
\left( z{\d\over\d z} z{d\over\d z}  +  z{d\over\d z}  + 2 \right) \chi_1 = 
- \left(2 +  z{\d\over\d z}\right) \chi_1^2,
\ee
with $\chi_1(\infty) = 0$.  Let's assume an expansion for $\chi_1$ of
the form given in equation (\ref{Eq:8}) with the substitution
$(\gamma_1, \gamma_2) \to (\delta_1,\delta_2)$.  The real parts of
$\delta_{1,2}$ must be negative so that the expansion has the right
limit at infinity, where $\chi_1\to 0$.  By construction we know that
$b_{00} = 0$.  In order for $\chi$ to be real, the matrix of
coefficients $b_{mn}$ must be Hermitian.  Now let's insert this ansatz
into our shifted differential equation [equation (\ref{Eq:9})].  For
convenience define $\Delta = m\delta_1+n\delta_2$, since then
\be
z {\d\over\d z}  z^{m\delta_1+n\delta_2} = z {\d\over\d z}  z^\Delta = \Delta \; z^\Delta,
\ee
and so equation (\ref{Eq:9}) becomes
\be
\left( \Delta^2 + \Delta + 2 \right) b_{mn} = 
-\left(\Delta+2\right) \sum_{i=0}^m \sum_{j=0}^n b_{ij} \; b_{m-i;n-j}.
\ee
This implies the recursion relation
\be
\label{E:recur2}
b_{mn} = 
-{\left(\Delta+2\right)\over \left( \Delta^2 + \Delta + 2 \right) }
 \sum_{i=0}^m \sum_{j=0}^n b_{ij} \; b_{m-i;n-j}.
\ee
Since $b_{00}=0$, the right hand side does not involve $b_{mn}$
itself.  Thus $b_{mn}$ can be calculated once the lower-order $b_{ij}$
are known. In particular once $b_{01}= b_{10}^*$ is specified, all
other coefficients can be calculated recursively. Unfortunately the
only known way of calculating $b_{01}= b_{10}^*$ is by performing a
numerical integration (\emph{e.g.}, Runge--Kutta) out to large radius
and then numerically fitting the data.  Summing up, we obtain
\be
\chi(z) = 1 + \sum_{m=0}^\infty \sum_{n=0}^\infty b_{mn} \; z^{m\delta_1+n\delta_2}.
\ee
A similar power series will exist for $\zeta(z)$. Indeed suppose
\be
\zeta(z) = 1 + \sum_{m=0}^\infty \sum_{n=0}^\infty a_{mn} \; z^{m\delta_1+n\delta_2}.
\ee
Then
\be
a_{mn} = (\Delta+1) \; b_{mn} = (m\delta_1+n\delta_2+1) \; b_{mn}.
\ee
In terms of the physical variables $m(r)$ and $p(r)$
\be
m(r) = 2 c_s^2 r \chi(r) 
= 2 c_s^2 r 
\left\{ 1 + \sum_{m=0}^\infty  \sum_{n=0}^\infty b_{mn} 
\left({4\pi p_0 r^2\over c_s^4}\right)^{(m\delta_1+n\delta_2)} \right\},
\ee
and
\be
p(r) = {\zeta(z) c_s^4\over 2 \pi r^2} 
=   
{c_s^4\over 2 \pi r^2} 
\left\{ 1 + \sum_{m=0}^\infty  \sum_{n=0}^\infty a_{mn} 
\left({4\pi p_0 r^2\over c_s^4}\right)^{(m\delta_1+n\delta_2)} \right\}.
\ee
These are Frobenius-like power series with fractional complex
exponents for the isothermal star [recall $\delta_{1,2} = (-1\pm
i\sqrt{7})/2$], where the coefficients $a_{mn}$ and $b_{mn}$ are
specified by the recursion relations determined above. These
expansions are considerably more general than the discussion presented
in Chandrasekhar~\cite{Chandrasekhar}, and the related results for the
relativistic case implicit in Harrison \emph{et al.}\cite{Harrison}.
Those results are tantamount to just keeping the first sub-asymptotic
term, which is also contained in our solution, namely
\be
m(r) 
\approx 2 c_s^2 r 
\left\{ 
1 +  2 \Re \left[ b_{10} \left({4\pi p_0 r^2\over c_s^4}\right)^{\delta_1} \right] 
\right\},
\ee
and
\be
p(r) \approx  
{c_s^4\over 2 \pi r^2} 
\left\{ 
1 + 2 \Re \left[ a_{10} \left({4\pi p_0 r^2\over c_s^4}\right)^{\delta_1} \right] 
\right\}.
\ee
Using analytic techniques, the series can be shown to converge for
$|z| \geq {121\over4}|b_{01}|^2$. This is a weak bound but at least
establishes a non-zero radius of convergence.  Numerically fitting
$|b_{01}|$ by matching to a 4th-order Runge--Kutta integration implies
guaranteed convergence for $|z|>5.246853637$.  Direct numerical
evidence suggests that the power series is reasonably well behaved
down to $z\approx 3$ with the first three terms providing $0.5\%$
accuracy.  By numerically applying the ratio test, we find that this
power series seems to converge for $ |z| \geq 0.165827$.  Indeed if we
keep the first $100$ ``levels'' ($n+m\leq100$) then the power series
reproduces a Runge-Kutta integration to within 6 significant figures
for $z>0.21$.

The combination of both power series, a Taylor series around $r=0$ and
a fractional power series around $r=\infty$, provides a solution
everywhere to arbitrarily good accuracy, since the two expansions have
overlapping radii of convergence.
%
\subsection{Power series in terms of the pressure}
\subsubsection{Linearization about the fixed points}
%
Instead of expanding $\chi$ and $\zeta$ in terms of radius $r$, we now
want to develop an expansion in terms of pressure $p$. This will be
useful if we want to generalize beyond a simple linear equation of
state.

We can rewrite the equation of hydrostatic equilibrium in the
following manner:
\be
{\d p\over\d r} = {2 p(r) \chi(r)\over r}.
\ee
Now, by applying the chain rule to $\d \zeta/\d p$, differentiating
equation (\ref{zeta:1}) with respect to $r$, and plugging in our new
expression for the equation of hydrostatic equilibrium, we obtain
\be
\label{zeta:2}
p\; {\d \zeta\over\d p} = \zeta - {\zeta\over \chi}.
\ee
The same can be done to equation (\ref{chi:1}) to obtain
\be
\label{chi:2}
p\; {\d \chi\over\d p} = {1 \over 2} \left(1-{\zeta\over \chi}\right).
\ee
This new system of differential equations in terms of the pressure
still has a stagnation point at $(\zeta,\chi)$ = $(1,1)$. This is the
same stagnation point that we encountered earlier. This stagnation
point is again an attractor, which means that for any initial
condition as $p\to 0$ almost all solutions of the system approach this
solution. Note that the second stagnation point (the repeller) at
$(0,0)$ is still there but somewhat disguised, since the system in
terms of $p$ is highly singular there. We will not need the second
stagnation point in the following discussion.  Linearizing about the
$(1,1)$ stagnation point we obtain
\be
p\; {\d \zeta_1\over\d p} = \chi_1;
\ee
\be
p\; {\d \chi_1\over\d p} = {1 \over 2} (\chi_1-\zeta_1).
\ee
Close to the $(1,1)$ stagnation point, the solutions to the system are of 
the form
\be
{\hat\zeta_1(p)\choose\hat\chi_1(p)} = 
\exp\left\{ \left[ \begin{array}{cc}0\;&1\\-1/2\;&1/2\end{array} \right] 
\ln(p/p_i) \right\}
\;{\hat\zeta_1(p_i)\choose\hat\chi_1(p_i)}. 
\ee
The eigenvalues of this matrix are $\lambda = (1\pm i \sqrt{7})/4$.
Note that this is simply $-\delta/2$, where $\delta$ are the exponents
as encountered when working with the variable $r$.

\subsubsection{Power series about $(1,1)$ in terms of the pressure}
%
It is essential to decouple the system of differential equations in
order to obtain a power series expression for the mass and the radius
in terms of the pressure That is, we want to replace the first-order
system by a single second-order equation.  In order to decouple
equations (\ref{zeta:2}) and (\ref{chi:2}), we solve for $\zeta$ in
terms of $\chi$ from equation (\ref{zeta:2}):
\be
\zeta = \chi \left(1- 2 p {\d \chi\over\d p} \right).
\ee 
Now, if we insert this expression back into the first differential
equation, we obtain
\be
-1 + \chi + p{\d \chi\over\d p} = 
2 \left\{
\chi \;p{\d \chi\over\d p}\; - 
\left(p \;{\d \chi\over\d p}\right)^2 -
\chi \; p {\d \over\d p}\; p {\d \over\d p}\; \chi
\right\}.
\ee
We can see from the analysis in the previous section that
$\chi(r=\infty) = \chi(p=0) = 1$.
It is convenient to write $\chi=1+\hat\chi$ because we now have
$\hat\chi(p=0) = 0$.
After this shift in variables, our decoupled differential equation becomes
\be
\label{shifted:1}
\left\{ 1 - p{\d\over\d p} + 2 p {\d \over\d p}\; p {\d \over\d p}\right\} \hat\chi = 
\left\{
p{\d \hat\chi\over\d p}\; - 
p {\d \over\d p}\; p {\d \over\d p}\right\} \hat\chi^2.
\ee
Now, in line with Theorem \ref{theo:5} and our general discussion
regarding Frobenius-like series, we assume an expansion of the form
\be
\hat\chi(p) = \sum_{m=0}^\infty \sum_{n=0}^\infty d_{mn} \; 
\left({p \over p_0}\right)^{m\lambda_1+n\lambda_2}.
\ee
Previously we used the dimensionless variable $z = ({\sqrt {4 \pi p_0}
/ c_s^2}) r$ to create a power series expansion about the stagnation
point $(1,1)$ for $\chi$ and $\zeta$. This time we find it more
convenient to expand in powers of $p/p_0$. 
It is interesting to note that our eigenvalues, $\lambda_{1,2}$, make physical sense
because we need positive real parts in order to satisfy the boundary
condition $\chi(0)=0$.  For $\hat\chi$ we have $d_{00} = 0$ and the
Hermiticity condition discussed previously still holds.
Again, let us for simplicity define $\Lambda = m\lambda_1 + n\lambda_2$.
We then have
\be
p {\d\over\d p}  \left({p\over p_0}\right)^{m\lambda_1+n\lambda_2} = p {\d\over\d p}
\left({p\over p_0}\right)^\Lambda = \Lambda \left({p\over p_0}\right)^\Lambda.
\ee
And therefore equation (\ref{shifted:1}) implies
\be
d_{mn} =  
{\Lambda \left(1- \Lambda\right)
\over
{\left( 1-\Lambda+2\Lambda^2\right)}}  
\sum_{i=0}^m \sum_{j=0}^n d_{ij} \; d_{m-i;n-j}.
\ee
Note that by construction $d_{00} =0$, so once $d_{01}=d_{10}^*$ is
specified all other coefficients can be calculated recursively.  Now
this recursion relation holds for the coefficients in $\hat\chi$
only. For $\chi$ itself we have
\be
\chi(p/p_0) = 
1 + \sum_{m=0}^\infty \sum_{n=0}^\infty d_{mn} \; 
\left({p \over p_0}\right)^{m\lambda_1+n\lambda_2}.
\ee
A similar power series will exist for $\hat\zeta({p/p_0})=
\zeta({p/p_0})-1$. Let us assume the following expansion
\be 
\hat\zeta({p/ p_0})= \sum_{m=0}^\infty \sum_{n=0}^\infty c_{mn} \left({p \over p_0} \right)^\Lambda.
\ee
Solving for $\zeta$, from equation (\ref{zeta:2}) we obtain
\be
\zeta = \chi - p {\d \chi^2\over\d p},
\ee
which, upon shifting to the fixed point, is equal to 
\be
\hat\zeta = \hat\chi - p {\d \over\d p} \left(2\hat\chi+\hat\chi^2 \right).
\ee
Comparing exponents we obtain
\be
c_{mn} =  d_{mn} - \Lambda \left(2  d_{mn} + 
\sum_{i=0}^m \sum_{j=0}^n d_{ij} d_{m-i;n-j} \right).
\ee
Using the recursion relation for the $d_{mn}$ 
\be
c_{mn} =  d_{mn} - \Lambda \left(2 d_{mn} +
{\left( 2 \Lambda^2 - \Lambda +1 \right)
\over \Lambda \left(1-\Lambda\right)}  d_{mn} \right) = -{2\Lambda\over1-\Lambda} \; d_{mn}.
\ee
Note that by construction $c_{00} =0$.  Now this relation holds for
the coefficients in $\hat\zeta$ only. For $\zeta$ itself we have
\be
\zeta(p/p_0) = 
1 + \sum_{m=0}^\infty \sum_{n=0}^\infty c_{mn} \; 
\left({p \over p_0}\right)^{m\lambda_1+n\lambda_2}.
\ee
We could of course unwrap the definitions of $\zeta$ and $\chi$ to
obtain Frobenius-like power series for the physical observables $r(p)$
and $m(p)$.
%
\subsection{Mass, pressure, and radius oscillations of a star 
on the verge of collapse ($p_0 \to \infty$)}
%
Let us now consider a more realistic equation of state:
\be
\rho(p) = \left\{ 
          \begin{array}{l@{\quad:\quad}l}
                       \rho_c= p_c/c_s^2 & p<p_c 
                        \\ 
                         {p / c_s^2} & p>p_c 
          \end{array} 
          \right.\, ,
\ee
where $p_c$ and $\rho_c$ stand for the pressure and density at the
surface of the core.  Most of the analysis carried until now still
holds, at least for the central region of the star. But we now chose
to replace the previous equation of state [$\rho(p) = p/c_s^2$, which
was linear out to arbitrarily low density] with a new equation that is
somewhat more realistic. We have a dense core described by a linear
equation of state surrounded by an envelope of constant density in
which the pressure drops rapidly to zero.

Since in the previous section we have established the existence of a
power series for $\zeta(p/p_0)$, we can formally solve equation
(\ref{zeta:3}) for the radius of the core itself by setting $p \to
p_c$ (at this stage making no comment about the envelope). Hence,
\be
r_\mathrm{core}(p_c,p_0)
=
{\sqrt{c_s^4\over{2\pi p_c}} \sqrt{\zeta(p_c/p_0)}} 
=
{c_s^2 \over{\sqrt{2 \pi p_c }}} 
\sqrt{ 1 + 
\sum_{m=0}^\infty \sum_{n=0}^\infty
{c_{mn}} \; \left({p_c \over p_0} \right)^{\Lambda} }.
\ee
This last equation clearly shows the oscillatory behaviour of the core
radius as a function of central pressure, due to the imaginary
exponents in the power series.

If we now look at equation (\ref{chi:3}) and solve for the mass of the core
in terms of the compactness and radius we obtain
\begin{eqnarray}
m_\mathrm{core}(p_c,p_0)
&=&
2c_s^2 \; r_\mathrm{core}(p_c,p_0) \; \chi(p_c/p_0)
\\
&=&
{2c_s^4\over{\sqrt{2 \pi p_c}}} 
\;
\sqrt{ 1 + 
\sum_{m=0}^\infty \sum_{n=0}^\infty
{c_{mn}} \; \left({p_c \over p_0} \right)^{\Lambda} }
\;
\left\{ 
1 + 
\sum_{m=0}^\infty \sum_{n=0}^\infty
{d_{mn}} \; \left({p_c \over p_0} \right)^{\Lambda} 
\right\}.
\nonumber
\end{eqnarray}
As we can see from this equation the mass of the core will also
oscillate as a function of central pressure due to the imaginary part
of the exponent of the power series. Indeed there is a related power
series such that
\be
m_\mathrm{core}(p_c,p_0)
=
{2c_s^4\over{\sqrt{2 \pi p_c}}} 
\;
\left\{ 
1 + 
\sum_{m=0}^\infty \sum_{n=0}^\infty
{m_{mn}} \; \left({p_c \over p_0} \right)^{\Lambda} 
\right\},
\nonumber
\ee
with the Hermitian coefficients $m_{mn}$ being in principle calculable
in terms of the $c_{mn}$ and $d_{mn}$.

The above expression gives us the mass of the core of the star, not of
the star itself.  In order to calculate the total mass of the star we
need to first calculate the mass of the envelope of the star:
\be
m_\mathrm{envelope}(r)= \int _{r_\mathrm{core}}^{r}{4 \pi r'^2 \rho_c}\,\d{r'}.
\ee
Hence,
\be
m_\mathrm{total}(r) = {4\over 3} \pi \rho_c (r^3-r_\mathrm{core}^3) + m_\mathrm{core}.
\ee
We calculate $r_\mathrm{surface}$
from our equation of state and equation of hydrostatic equilibrium:
\be
p_\mathrm{envelope}(r) = 
p_c - \int_{r_\mathrm{core}}^{r} {\rho_c \; m_\mathrm{envelope}(r') \over{r'^2}}\,\d{r'},
\ee
where we have used the fact that $\rho \to \rho_c$ is constant
throughout the envelope. Integrating
\be
p_\mathrm{envelope}(r) =  p_\mathrm{c} + 
\left( \rho_c m_\mathrm{core} - {4 \over 3} \pi \rho_c^2 r_\mathrm{core}^3 \right)
\left( {1 \over r} - {1 \over r_\mathrm{core}} \right) + {2 \over 3} \pi \rho_c^2
\left( r_\mathrm{core}^2 - r^2 \right).
\ee
We can solve for $r_\mathrm{surface}$ by solving for
$p_\mathrm{envelope}(r=r_\mathrm{surface})=0$. The resulting cubic equation is
\be
p_\mathrm{envelope}(r=r_\mathrm{surface})=0=C_1 \; r_\mathrm{surface}^3 + C_2 \; r_\mathrm{surface} - C_3,
\ee
where we have chosen the constants $C_1$, $C_2$ and $C_3$ to be
\be
C_1 = {2 \over 3} \pi \rho_c^2,
\ee
\be
C_2 = {\rho_c m_\mathrm{core} \over r_\mathrm{core}} - 2 \pi \rho_c^2 r_\mathrm{core}^2 - p_\mathrm{c},
\ee
\be
C_3 = \rho_c m_\mathrm{core}- {4 \over 3} \pi \rho_c^2 r_\mathrm{core}^3.
\ee    
This cubic equation can be explicitly [if tediously] solved for
$r_\mathrm{surface}$ as a function of $\rho_c$, $m_{core}$, and
$r_{core}$. Suffice it to say that we now have implicitly solved for
the total mass of the star ($M_\mathrm{total}$), where we have taken
$r_\mathrm{surface}$ as a parameter (which can ultimately be
expressed, via $m_{core}$, and $r_{core}$, as an oscillating function
of $p_c$ and $p_0$). That is 
\be 
M_\mathrm{total}(p_c,p_0)=
{4\over 3} \pi \rho_0 (r_\mathrm{surface}(p_c,p_0)^3
-r_\mathrm{core}(p_c,p_0)^3)
+ m_\mathrm{core}(p_c,p_0).
\ee 
Ultimately, since the individual terms above exhibit oscillatory
behaviour, there will be some Frobenius-like power series such that
\be 
M_\mathrm{total}(p_c,p_0) =
M_\mathrm{total}(p_c,p_0\to\infty) \; 
\left\{ 1 + \sum_{m=0}^\infty
  \sum_{n=0}^\infty {M_{mn}} \; \left({p_c \over p_0}
  \right)^{\Lambda} \right\}.  
\ee 
{From} this expression for the total mass of the star we can clearly
see the oscillatory behaviour mentioned earlier.

Note that this behaviour is much more extensive than the scaling
results discussed in Chandrasekhar~\cite{Chandrasekhar} [see also
Harrison \emph{et al}\cite{Harrison}]. Their result is equivalent to
keeping only the first nontrivial term in the power series. That is:
\be
M_\mathrm{total}(p_c,p_0)
\approx
M_\mathrm{total}(p_c,p_0\to\infty) 
\;
\left\{ 
1 + 2 \Re\left[ M_{10} \left({p_c \over p_0} \right)^{\delta_1}  \right]
\right\}.
\ee
We now see that the scaling behaviour they discussed is only the first
term in an infinite series.

It may be objected that our current analysis is only valid for the
special case of a linear equation of state with a constant density
envelope. However, from the way we have set up the general theory it
is clear that this restriction is not particularly serious. First, any
reasonable equation of state will saturate to a linear at high enough
densities, so our results can be applied without modification to the
core of any reasonable stellar configuration as the central pressure
$p_0\to\infty$. Furthermore, suppose one has some generic equation of
state. Then the exact system will not be scale invariant. Nevertheless
there will be a special critical configuration in which the central
pressure is infinite, and since by hypothesis the equation of state
becomes linear at sufficiently high pressure, the critical solution
will exhibit approximate scale invariance in the high-pressure region
--- and this will correspond to a fixed point in the autonomous
system. Linearization around the fixed point will generate some
critical exponents $(\delta_1 ,\delta_2)$, which depend only on the
high-pressure limit of the equation of state.  But deviations from the
scale invariant critical configuration in the core of the star will
now feed into the exact scale non-invariant equations, still leading
to the total mass being given in terms of some Frobenius-like series.
Ultimately this Frobenius-like power series behaviour is generic to
many systems of differential equations and intrinsically does not have
anything to do with gravity {\emph{per se}}.
%
\section{Application: Isothermal relativistic stars}
\label{C:tov}
\subsection{Background behaviour}
In order to treat relativistic stars, we need to consider the
relativistic equation of hydrostatic equilibrium (TOV equation)
\be
\label{tov}
{\d p\over\d r} = - \left(\rho + {p\over c^2}\right) 
{\left(m + 4 \pi r^3 {p/c^2} \right)\over{r^2\left(1-{2m/ (r c^2)}\right)}}.
\ee
We again have a system of two first-order differential equations,
namely equations (\ref{tov}) and the mass equation (\ref{mass}), which
must be supplemented by an algebraic equation of state.  Note that in
the TOV equation $c$ is the speed of light, and that as $c\to \infty$
we formally recover Newtonian physics. We will consider an equation of
state identical to that used in the previous chapter, \emph{i.e.}
$p(r) = c_s^2 \; \rho(r)$, where $c_s$ stands for the speed of sound
in the fluid.  Making the appropriate substitutions we obtain the
closed system
\bd
S \quad 
\left\{ \begin{array}{l}
        {\displaystyle \vphantom{\Bigg|}
         {\d p\over\d r} = -p \left({1\over c_s^2}+{1\over c^2}\right) 
                          {\left(m + {4 \pi r^3 p/ c^2} \right)\over{r^2\left(1-{2m/ (r c^2)}\right)}} },      
        \\
        \\
        {\displaystyle \vphantom{\Bigg|} 
         {\d m\over\d r} = {4 \pi r^2 p\over c_s^2}} .
        \end{array} 
\right.
\ed 
Note that (as in the Newtonian case) this system $S$ is 
invariant under the substitutions $r \to ar$, $p \to a^{-2} p$ and
$m\to a\;m$.  Revisiting our theoretical analysis we also discover
that, $(y_1,y_2) = (p,m)$ and $(p_1, p_2)=(-2,1)$, again as in the
Newtonian case.

It is possible to systematically transform the system $S$ into an
equidimensional-in-$r$ system $S_1$, and subsequently into an
autonomous system $S_2$.  Let us first substitute $\zeta=2\pi p(r)
r^2/c_s^4$ and $\chi=m/(2r c_s^2)$. Let us also define $t=\ln(r)$ so
$\d/\d t = r \d/\d r$.  Then, the system becomes
\bd
S_1 \left\{ \begin{array}{l}
            \displaystyle
            {\d \zeta\over\d t} = 2\zeta - 2 \zeta \left(1+\beta^2\right) {\left(\chi + \beta^2 \zeta \right)\over{1-4 \beta^2\chi}}, 
            \\
            \\
            \displaystyle
            {\d \chi\over\d t} = \zeta - \chi,
           \end{array} \right.
\ed
where $\beta = {c_s/ c}$.  As we can clearly see $S_1$ is now
autonomous, and therefore, translation invariant.  There are two fixed
points, namely $\left(\alpha,\alpha\right)$ and $ \left(0,0\right)$,
where $\alpha = 1$/$(1+ 6 \beta^2 + \beta^4)$. Note that $\beta\to0$
corresponds to the Newtonian limit.

As we found previously, the first fixed point is an attractor as
$t\to\infty$, which corresponds to $1/r^2$ behaviour for the pressure.
The second fixed point at $(0,0)$ is a repeller, as we also found in
the Newtonian case. This fixed point corresponds to regular $r^2$
behaviour for $\zeta$ and $\chi$ at the center of the star, plus an
irregular and unphysical behaviour for $\chi$ that goes as $1/r$.  The
center of the star is at $r=0$ ($t=-\infty$). For a regular star with
finite $p_0$ we have $\zeta=0$ and $\chi=0$ as $t\to-\infty$. So
solving $S_1$ is equivalent to looking for the unique integral curve
that emerges from $(0,0)$ and terminates at $(\alpha,\alpha)$. Since
system $S$ is scale invariant and its associated autonomous system
$S_1$ has fixed points, we can apply Theorem \ref{Thm:1} and easily
check that the background solution to $S$ is
\be
p(r) =  C_1 \; r^{-2},
\ee
and
\be
m(r) =  C_2 \; r^1,
\ee
where $C_1$ and $C_2$ are constants. Inserting back into $S$ and
solving for the constants we see that the $(\alpha,\alpha)$ fixed
point corresponds to the power-law solution
\be
\label{Eq:5b}
m(r) = {2c_s^2\over {1+6\beta^2+\beta^4}}\; r,
\ee
and 
\be
\label{Eq:6b}
p(r) = {c_s^4\over {2 \pi} r^2} \; {1\over {1+6\beta^2+\beta^4}}.
\ee
Note that if we take the limit $c \to \infty$, we recover the exact
solutions obtained in the previous chapter for the Newtonian case.
Also note that the background solution for the $(0,0)$ fixed point
corresponds to the trivial $p(r)=0=m(r)$.
%
\subsection{Power series in terms of the radius}
\subsubsection{Linearization about the fixed points}
We suspect that, as in the Newtonian case, the pressure and the mass
will have a series expansion in the form of an iterated power law. In
order to determine the exponents of the power, let's analyze each
fixed point in more detail. Linearizing about the $(\alpha,\alpha)$
fixed point we obtain the following system
\bd
S^{linear}_1 \left\{ \begin{array}{l}
            \displaystyle {\d \hat\zeta\over\d t} = A  \hat\zeta + B \hat\chi, 
            \\
            \\
            \displaystyle {\d \hat\chi\over\d t} = \hat\zeta - \hat\chi,
           \end{array} \right.
\ed
where
\bd
A = {-2 \beta^2 \over{1+\beta^2}} ; 
\qquad  
B = {-2 \left(1+ 5 \beta^2 \right)\over{\left(1+\beta^2\right)^2}},
\ed
and $\hat\zeta$ and $\hat\chi$ are small perturbations.  Therefore,
close to the stagnation point, solutions are of the form
\be
{\zeta(t)\choose\chi(t)} = 
\exp\left\{ 
\left[ \begin{array}{cc}A&B\\1&-1\end{array} \right] (t-t_i) 
\right\}
\; {\zeta(t_i)\choose\chi(t_i)},
\ee
The exponents of the power law are given by the eigenvalues of this
matrix, which are
\bd
\delta_{\pm} = 
{-1-3 \beta^2 \pm \sqrt{-7-42\beta^2+\beta^4}\over{2(1+\beta^2)}},
\ed
and the approximate linearized solutions approach the critical point
exponentially in $t$ --- as $\exp(\delta t) = r^\delta$.  These are
the exact same power exponents found by Harrison
\emph{et. at.}\cite{Harrison} Note that if we take the Newtonian limit
$c \to \infty$, \ie, $\alpha \to 1$ and $\beta \to 0$, we recover the
eigenvalues for the Newtonian case.

Let us now analyze the $(0,0)$ fixed point. Making the substitution
$\chi = \hat\chi$ and $\zeta = \hat\zeta$, where $\hat\chi$ and
$\hat\zeta$ are to be viewed as small perturbations, we obtain the
following system
\bd
S^{linear}_2 \left\{ \begin{array}{l}
            \displaystyle {\d \hat\zeta\over\d t} =  2 \hat\chi, 
            \\
            \\
            \displaystyle {\d \hat\chi\over\d t} = \hat\zeta - \hat\chi.
           \end{array} \right.
\ed
Close to the stagnation point, solutions to this system have the form
\be
{\zeta(t)\choose\chi(t)} = 
\exp\left\{ \left[ \begin{array}{cc}0&2\\1&-1\end{array} \right] (t-t_i) \right\}
\; {\zeta(t_i)\choose\chi(t_i)}.
\ee
This is the second unstable stagnation point, which was mentioned in
the previous section. This fixed point is a repeller with eigenvalues
$\left(\gamma_1,\gamma_2\right) = (2,-1)$. Note that these are the same eigenvalues
found for the fixed point $(0,0)$ in the Newtonian analysis.
We have now determined the exponents of the critical power-law
behaviour for both fixed points. 
%
\subsubsection{Power series about $(0,0)$ in terms of the radius}
%
Based on Theorem \ref{theo:5} and our general argument for the
existence of Frobenius-like series let us now assume an expansion for
$\chi$ and $\zeta$ of the form
\be
\label{series}
\chi(r) = \sum_{m=0}^\infty \sum_{n=0}^\infty b_{mn} \; r^{m\gamma_1+n\gamma_2}; \qquad
\zeta(r) = \sum_{m=0}^\infty \sum_{n=0}^\infty a_{mn} \; r^{m\gamma_1+n\gamma_2}.
\ee
We now need to calculate the actual coefficients $b_{n}$ and
$a_{n}$ of the series. In order to find them, we need to first
decouple the system $S_1$.  Solving for $\zeta$ in the $d\chi/dr$
equation in $S_1$ and inserting this into the $d\zeta/dr$ equation we
obtain
\bd
\label{decoupled}
r {\d \over\d r} r{\d \over\d r}\chi - 2\chi - r{\d \over\d r} \chi  = 
- 2 {\left(1+\beta^2\right)\over {1-4\beta^2\chi}}
\left(\chi + r{\d \over\d r} \chi\right) \left[\chi\left(1+\beta^2\right)+\beta^2 r{\d \over\d r}\chi\right].
\ed
It is now convenient to use the dimensionless variable $z = r
\sqrt{4\pi p_0}/c_s^2$. Rearranging the above, we obtain
\bea
&&\left[2 + z{\d \over\d z} - z {\d \over\d z} z{\d \over\d z}\right]\chi 
\\
&&\qquad= \left[ 2 \left(1+6\beta^2 +\beta^4\right) +  
 \left(1+5\beta^2+2\beta^4\right) z{\d \over\d z} 
- 2\beta^2 z{\d \over\d z} z{\d \over\d z} \right]\chi^2  
\nonumber
\\ 
&&
 \qquad\qquad 
+
2\beta^2\left(3+ \beta^2\right) \left(z {\d \over\d z} \chi\right)^2. 
\nonumber
\eea
Note here that $\chi(\infty)=\alpha$ and that if we take $\beta \to 0$
we recover the Newtonian result $\chi(\infty)=1$. We now have
a second order decoupled differential equation for $\chi$. 
Before solving for the coefficients of the series, we remind ourselves that
for this case the $1/r$ behaviour is unphysical. Hence, let us
suppress all occurrences of $\gamma_2=-1$. and restrict our analysis to
$\gamma_1=2$. Instead of equation (\ref{series}) we now have
\ba
\label{E:series2}
&& \chi(z) = \sum_{n=1}^\infty b_{n} \; z^{2n} = 
\sum_{n=1}^\infty b_{n} \; \left({4\pi p_0 r^2\over c_s^4}\right)^n; \\
\label{E:series22}
&& \zeta(z) = \sum_{n=1}^\infty a_{n} \; z^{2n} = 
\sum_{n=1}^\infty a_{n} \; \left({4\pi p_0 r^2\over c_s^4}\right)^n.
\ea
Plugging this ansatz into our decoupled differential equation, we
obtain the following recursion relation for the coefficients $b_{n}$
\bea
b_n &=&
- 
{
{\left(1+6\beta^2+\beta^4\right) 
+\left(1+5\beta^2+2\beta^4\right)n - 4 \beta^2 n^2}
\over
{\left(n-1\right)\left(2n+1\right)}
} 
\;\;\sum_{i=1}^{n-1} b_i b_{n-i} 
\nonumber\\
&& -
{4\beta^2\left(3+\beta^2\right)\over{\left(n-1\right)\left(2n+1\right)}} 
\;\;\sum_{i=1}^{n-1} i \left(n-i\right) b_i \; b_{n-i}.  
\eea
With this recursion relation, we have an expression for $\chi(z)$
equation (\ref{E:series2}) in the form of a power series close to
the fixed point $(0,0)$.

A similar power series will exist for $\zeta$. Using the $\d m/\d r$
equation from $S_1$, we observe that (as in the Newtonian case)
\be
a_{n} = (2n+1) \; b_{n}. 
\ee
In order to completely specify the power series, we need to have
boundary conditions. The boundary conditions 
to begin the recursion relation are $b_0=0$ and $b_1=1/6$, which
are the same conditions used previously in the Newtonian analysis of
the system. Choosing $b_1=1/6$ is equivalent to the specific choice
of $z$ made above. Hence, inserting these conditions into the $a_n$
equation we obtain $a_0=0$ and $a_1=1/2$. We have now a complete
expression for $\chi$ and $\zeta$ close to the $(0,0)$ stagnation
point given by equation (\ref{E:series2}) and (\ref{E:series22}).

The radius of convergence $R(\beta)$ of the $\chi$ series will be the
same as that for the $\zeta$ series, and is now a function of the
parameter $\beta$. We performed numerical calculations and discovered
that there is a finite radius of convergence for almost all values of
$\beta$. In order to better understand the convergence of the series,
let us define two regions: Region I, ranging from $0<\beta<0.0722$ and
Region II, defined from $0.0722<\beta<1$.

The radius of convergence is well-behaved as a function of $\beta$ in region I,
allowing us to recover the Newtonian upper bound at $\beta=0$ ($R(0)\approx3.2$.)
Our results also showed that the radius of convergence strictly increases
monotonically until it reaches $\beta \approx 0.0722$ at which 
$R(0.0722) \approx 3.6736$.
 
In region II, the global behaviour of $\beta$ is given by a power law
(approximately $R(\beta)\propto\beta^{-1}$), decreasing from $R
\approx 3.6736$ at $\beta \approx 0.0722$ to $R \approx 1.1218$ at
$\beta = 1$.  The local behaviour as a function of $\beta$ in region
II, however, is more complex, since there are values of $\beta$ for
which the radius of convergence drops to zero. The initial value at
which our series diverges is $\beta_0=0.0724$, and after that it will
diverge at $\beta=\beta_0+\tau n$, where $n$ is an integer
($n=0,1,2,3,\ldots$) and $\tau \approx 0.00026$.  It seems possible to
approximate the behaviour of the radius of convergence in region II
with a damped oscillating function,
\emph{e.g.} $R(\beta) \propto \cos(2\pi\beta/\tau+\phi_0)/\beta$. 
%
\subsubsection{Power series about ($\alpha,\alpha$) in terms of the radius}
%
In order to determine the coefficients $b_{mn}$ and $a_{mn}$ for the
power series about the fixed point $(\alpha,\alpha)$ we need to make a
shift to $\hat\chi = \chi - \alpha$ in our decoupled differential
equation (\ref{decoupled}). By doing so, equation (\ref{decoupled})
can be simplified and put in operator form to obtain
\begin{eqnarray}
&&\left[
\left(1-2\alpha  \left[1+5\beta^2+2\beta^4\right]\right) z{\d \over\d z} 
- 
\left(1-4\alpha\beta^2\right) z {\d \over\d z} z{\d \over\d z}  
- 
2 
\right]\hat\chi 
\nonumber\\
&& 
\qquad\qquad =
\left[ 
2 \left(1+6\beta^2 +\beta^4\right) 
+  
\left(1+5\beta^2+2\beta^4\right) z{\d \over\d z} 
- 
2\beta^2 z{\d \over\d z} z{\d \over\d z} 
\right]\hat\chi^2 
\nonumber\\
&& 
\qquad\qquad\qquad
+ 2\beta^2\left(3 + \beta^2\right) \left(z {\d \over\d z}\hat\chi\right)^2. 
\end{eqnarray}
Again, we have a second order decoupled differential equation for
$\hat\chi$. Note that for our series expansion we have the expression given in equation
(\ref{series}), where we must take $\lambda_{1,2} \to \delta_{1,2}$.
Plugging this ansatz into our decoupled differential
equation, we obtain a recursion relation for the coefficients $b_{mn}$
\begin{eqnarray}
b_{mn} 
&=& -{
{2\left(1+6\beta^2 +\beta^2\right) +
\left(1+5\beta^2+2\beta^4\right)\Delta
-2\beta^2\Delta^2}
\over
{\left(1-4\beta^2\alpha\right)\Delta^2
-
\left(1-2\alpha  \left[1+5\beta^2+2\beta^4\right]\right)\Delta
+2}
} 
\;\;
\sum_{j=0}^{m}\sum_{i=0}^{n} b_{ij} b_{m-i;n-j}
\nonumber\\ 
&&  
-{2\beta^2\left(3+\beta^2\right)
\over
{\left(1-4\beta^2\alpha\right)\Delta^2
-
\left(1-2\alpha  \left[1+5\beta^2+2\beta^4\right]\right)\Delta
+2}} 
\nonumber\\ 
&& \qquad\qquad \times
\sum_{j=0}^{m}\sum_{i=0}^{n} ij 
\left(m-j\right)\left(n-i\right) b_{ij} b_{m-i;n-j}, 
\end{eqnarray}
where $\Delta=m\delta_1+n\delta_2$. With this recursion relation, we
again have a complete expression for $\chi$ in the form of a power series
close to the fixed point $(\alpha,\alpha)$. $\zeta$ will also have a
power series expansion, with coefficients given (as in the
Newtonian case) by
\be
a_{mn} = (\Delta+1) \; b_{mn}.
\ee
Once more, we need boundary conditions to begin the recursion
relation.  We know analytically that $b_{00}=0$ and $a_{00}=0$. If we
knew $b_{01}=b_{10}^*$ [or equivalently $a_{01}=a_{10}^* =
(1+\delta_2) b_{01}$] then we could recursively calculate the entire
series. As in the Newtonian case, there is no analytic theory for
these coefficients.

Though we do not have an analytic way of specifying $a_{01}$ or
$b_{01}$, we do have the powerful result that $\zeta$ and $\chi$ are
[close to the fixed point $(\alpha,\alpha)$] given by power series of
the form
\bd
\chi(r) = \alpha + \sum_{m=0}^\infty \sum_{n=0}^\infty b_{mn} \left({4\pi p_0 r\over c_s^2}\right)^{m\delta_1+n\delta_2},
\ed
and
\bd
\zeta(r) = \alpha + \sum_{m=0}^\infty \sum_{n=0}^\infty a_{mn} \left({4 \pi p_0 r\over c_s^2}\right)^{m\delta_1+n\delta_2}.
\ed

In terms of the dimensional physical variables $p(r)$ and $m(r)$ this
can explicitly be put in Frobenius-like form
\bd
m(r) = 
2 c_s^2 r \;
\left[
\alpha + \sum_{m=0}^\infty \sum_{n=0}^\infty b_{mn} \left({4\pi p_0 r\over c_s^2}\right)^{m\delta_1+n\delta_2}
\right],
\ed
and
\bd
p(r) = {c_s^4\over4\pi r^2} \; 
\left[
\alpha + \sum_{m=0}^\infty \sum_{n=0}^\infty a_{mn} \left({4 \pi p_0 r\over c_s^2}\right)^{m\delta_1+n\delta_2}
\right].
\ed
%
\subsection{Power series in terms of the pressure}
%
Though we will not go into any of the details, it is clear from the
existence of the power series $p(r)$, that the series can be
``reverted'' to provide a power series for $r(p)$. The key points are
that $p(r)$ is monotonic, so the inverse $r(p)$ exists. Then from the
power series above we deduce the existence of a reverted
Frobenius-like power series
\be
r(p,p_0) = {c_s^2\over\sqrt{4\pi p}} \; 
\left[
\alpha + \sum_{m=0}^\infty \sum_{n=0}^\infty 
r_{mn}(\beta) \;\; \left({p\over p_0}\right)^{m\delta_1+n\delta_2}
\right].
\ee
From this we deduce that for a cutoff equation of state (linear in the
core, constant density in the envelope) even in the relativistic case
the core mass and core radius will exhibit damped oscillations as a
function of central pressure $p_0$.
\be
r_\mathrm{core} (p_c,p_0) = r(p=p_c,p_0) = {c_s^2\over\sqrt{4\pi p_c}} \; 
\left[
\alpha + \sum_{m=0}^\infty \sum_{n=0}^\infty 
r_{mn}(\beta) \;\; \left({p_c\over p_0}\right)^{m\delta_1+n\delta_2}
\right].
\ee
Similarly the total mass and total radius will exhibit
damped oscillations.

It should also be clear that the details of the equation of state at
low density do not matter. As long as the equation of state at high
densities is linear, then the analysis above implies that as the
central pressure goes to infinity the core radius and core mass will
undergo damped oscillations as a function of central pressure. Once
the core undergoes these oscillations, the envelope serves only to
communicate this internal behaviour out to the stellar surface. Thus
the total mass and total radius will similarly undergo damped
oscillations. These damped oscillations will involve an infinite power
series in terms of $p_0^\Delta$, which will contain the first-order
term discussed in Harrison~\emph{et al.}\cite{Harrison}, plus an
infinite collection of higher-order terms.

\section{Conclusions}
%
We have presented a unified formalism for calculating the solution to
scale invariant theories in the form of power laws, and deviations
from pure power-law behaviour.  We have proved that some of the
solutions to a scale-invariant differential system, whose associated
autonomous equation possesses fixed points, will exhibit pure
power-law behaviour.  We have also proved that after linearizing about
these fixed points, the solution to the scale-invariant linearized
system can also be written in the form of an independent collection of
power laws.  Generically, these collections of power-law solutions can
be combined to develop a Frobenius-like series in a neighborhood
surrounding the fixed point.

We applied these ideas to several problems taken from astrophysics,
such as static isothermal stars both in the Newtonian and TOV
frameworks.  We showed, both numerically and analytically, that in all
the static cases the solution to the scale-invariant differential
system associated with each example exhibits power-law behaviour as the
star approaches the point of collapse.  We also showed analytically
that the damped oscillations in the pressure, mass and radius of a
star, as the central density goes to infinity, can be represented as a
Frobenius-like power series, which is easily obtained by applying our
formalism.

More generally, it would be advantageous to consider truly dynamical
systems involving PDEs in both space and time. Work along these lines
is progressing.

%
\section*{Acknowledgments}
%
This research was partially supported by the US DOE.  Nicolas Yunes
wishes to acknowledge the support of Washington University through the
Mesmer and Washington University Scholarship, as well as support from
the Washington University Gravitation Group. We gratefully thank the
Washington University Gravitation Group for useful comments and
support.
%
\appendix
\section{Autonomous equation without fixed points}
\label{appe:1}
Caution must be exercised since not all scale invariant equations
possess fixed points. A scale invariant differential equation, whose
associated autonomous equation \emph{does not} have any fixed points,
will \emph{not} possess power-law solutions.  As an example, let's
analyze what happens to the following relatively simple differential
equation:
\be
\label{countereg}
F(x,y(x)) \equiv {\d y\over \d x} - {p \; y(x)\over x} - f\; x^{p-1} = 0,
\ee
where $f$ is a non-zero constant. This is easily verified to be a
scale-invariant equation and the associated equidimensional-in-$x$
equation is
\be
\widetilde F(x,w(x)) \equiv x^p \left( {\d w(x)\over\d x} -  {f\over x}\right) = 0.
\ee
This factorizes to give
\be
f(t) \; \bar F(z(t)) = e^{(p-1)t} \left( {\d z(t)\over\d t} - f \right) =0.
\ee
This last equation is autonomous and it does not possess a fixed point
since we are assuming that $f \not= 0$. The general solution for this
differential equation is easily verified to be
\be
y(x) = x^p \; \left(f \ln{x} + A \right).
\ee
Note that the solution to our example will be a power law if and only
if $f = 0$, regardless of the fact that our example is explicitly
scale invariant for all values of $f$.  In conclusion, scale
invariance by itself does not guarantee that the solution to a
scale invariant differential equation will exhibit power law
behaviour. More complicated examples of this behaviour can easily be
constructed, but this simple equation is already enough to make the
point that scale invariance does not necessarily lead to
power-law solutions.

However, if we impose \emph{both} scale invariance and the constraint
that the associated autonomous equation has fixed points, then the
solution will certainly exhibit power law behaviour.
%
\section{Limit cycles and discrete self-similarity}
\label{a:limit}

In the theory of autonomous differential equations the existence of a
fixed point is one of the simplest things one could wish for; the next
most complicated structure one might encounter is a limit cycle.

Suppose we start with a scale invariant differential equation and that
the associated autonomous equation has a limit cycle rather than a
fixed point. This means there is a special solution $z_*(t)$ with a
period $T$ such that
\be
z_*(t + n T) = z_*(t) \qquad \forall n \in Z.
\ee
In terms of the equidimensional variables this implies
\be
w_*( \exp(n T) x) = w_*(x).
\ee
That is, a limit cycle in the variable $t$ becomes a discrete
self-similarity in the variable $x$. Similarly in terms of the
original variable $y(x)$ we obtain discrete power-law behaviour
\be
y(x) = x^p \; w_*(x); \qquad \forall n\in Z: \quad y( \exp(n T) x) = \exp( p n T) \; y(x).
\ee
This is not the classic power-law behaviour discussed previously, but
is, in a sense, the next least complicated thing.
As an example, let's consider the second-order differential equation
\be
F(x,y(x)) \equiv  x^2 y''(x) - (2p-1) x y'(x) + p^2 y + \Omega^2 y(x) = 0
\ee
This is easily verified to be scale invariant with
index $p$. The associated equidimensional-in-$x$ equation is
\be
\widetilde F(x,w(x)) \equiv  x^2 w''(x) + x w'(x) + \Omega^2 w(x) = 0.
\ee
The associated autonomous equation is
\be
\bar F(z(t)) \equiv  \ddot z(t) + \Omega^2 z(t) = 0.
\ee
Now there is a trivial fixed point at $z=0$ but the general solution
to the autonomous equation is
\be
z(t) = A\cos(\Omega t) + B\sin(\Omega t).
\ee
In terms of the equidimensional variables
\be
w(x)  = \Re\left[ (A+i B) x^{i\Omega} \right].
\ee
Finally, in terms of $y(x)$
\be
y(x)  =  x^p \; \Re\left[ (A+i B) x^{i\Omega} \right].
\ee
This now is an explicit example of a discrete power law. The solution
is not scale invariant under arbitrary rescalings $x \to a\; x$, but
it is invariant under the specific rescaling $x \to x
\exp(2\pi/\Omega)$.

Note the implication: Discrete self-similarity and discrete power laws
are not as peculiar as one might at first imagine; instead they are
simply the next most complicated thing (after simple power-law
behaviour) that can happen in scale invariant systems. (Viewed in this
light the occurrence of discrete self-similarity in Choptuik's
critical solution~\cite{Choptuik,Gundlach} should not at all be
considered surprising.)
%
\section{Isothermal Lane--Emden equation.}
\label{a:lane-emden}

The traditional [isothermal] Lane--Emden equation is
\be
{1\over z^2} {\d\over\d z}\left( z^2 {\d \phi(z)\over\d z} \right) =  \exp(-\phi(z)).
\ee
\be
\phi(0) = 0;   \qquad \left.{\d \phi\over\d z}\right|_0 = 0.
\ee
This second-oder DE is completely equivalent to the first-order system
we derived for a Newtonian relativistic star. Indeed
\be
\phi(z) = - \ln\left[2\zeta(z)/z^2\right],
\ee
and the series we previously determined for $\zeta(z)$ in the
Newtonian case then carry over into series for the Lane-Emden function
$\phi(z)$. It is then straightforward to compute
\be
\phi(z) =
{1\over6}\,{z}^{2}-{\frac {1}{120}}\,{z}^{4}+{\frac {1}{1890}}\,{z}^{6}
-{\frac {61}{1632960}}\,{z}^{8}+{\frac {629}{224532000}}\,{z}^{10}
+O(z^{12}).
\ee
You can also derive this in a more direct way by putting a trial power
series directly into the Lane--Emden DE and equating coefficients.

Perhaps more surprising is the implication that at large $z$ the
Lane--Emden function should have an expansion of the form
\be
\phi(z) = \ln[z^2/2] + 
\sum_{m=0}^\infty  \sum_{n=0}^\infty q_{mn} \; z^{(m\delta_1+n\delta_2)} .
\ee
[Recall $\delta_{1,2} = (-1\pm i\sqrt{7})/2$.] The coefficients
$q_{mn}$ are in turn determined by the recursion relations for
$a_{mn}$ and $b_{mn}$ previously discussed. An alternative more direct
route is to write
\[
\phi(x) =  \ln[z^2/2] + q(z),
\]
and substitute into the Lane--Emden equation yielding
\be
\left\{ z {\d\over\d z} z{\d\over\d z} + z{\d\over\d z} \right\} q(z) = 
2 \left\{ \exp[-q(z)] - 1 \right\}.
\ee
Linearizing around the obvious solution $q(z)=0$ recovers the critical
indices $\delta_{1,2} = (-1\pm i\sqrt{7})/2$, while inserting a trial
series consisting of integer powers of these critical exponents will
determine all the higher coefficients $q_{mn}$ in terms of
$q_{01}=q_{10}^*$. Of course, as is usual, there is no analytic theory
for this first coefficient since it corresponds to finding the unique
boundary condition at infinity that leads to a regular star at the
center --- $q_{01}=q_{10}^*$ has to be determined numerically from an
outward integration of the Lane--Emden equation starting from the
origin where we do know the physical boundary conditions.

When it comes to writing down explicit recursion relations for the
coefficients, it is more convenient to work with the compactness
$\chi(z)$ in terms of which the [isothermal] Lane--Emden equation is
equivalent to equation (\ref{operator}) and explicit recursion relations
are given in equations (\ref{E:recur1}) and (\ref{E:recur2}).


%
\end{document}